# Ethical Decision Making During Automated Vehicle Crashes

Noah Goodall

## ABSTRACT

Automated vehicles have received much attention recently, particularly the DARPA Urban Challenge vehicles, Google's self-driving cars, and various others from auto manufacturers. These vehicles have the potential to significantly reduce crashes and improve roadway efficiency by automating the responsibilities of the driver. Still, automated vehicles are expected to crash occasionally, even when all sensors, vehicle control components, and algorithms function perfectly. If a human driver is unable to take control in time, a computer will be responsible for pre-crash behavior. Unlike other automated vehicles—such as aircraft, where every collision is catastrophic, and guided track systems, which can only avoid collisions in one dimension—automated roadway vehicles can predict various crash trajectory alternatives and select a path with the lowest damage or likelihood of collision. In some situations, the preferred path may be ambiguous. This study investigates automated vehicle crashing and concludes the following: (1) automated vehicles will almost certainly crash, (2) an automated vehicle's decisions preceding certain crashes will have a moral component, and (3) there is no obvious way to effectively encode complex human morals in software. A three-phase approach to developing ethical crashing algorithms is presented, consisting of a rational approach, an artificial intelligence approach, and a natural language requirement. The phases are theoretical, and should be implemented as the technology becomes available.

## INTRODUCTION

Automated vehicle technologies are the computer systems that assist human drivers by automating aspects of vehicle control. These technologies have a range of capabilities, from anti-lock brakes and forward collision warning, to adaptive cruise control and lane keeping, to fully automated driving. The National Highway Traffic Safety Administration (NHTSA) has defined five levels of vehicle automation, described in Table 1 *(1)*. The focus of this study is automation Levels 3 and 4, with some applications in Level 2. Throughout this paper, the term "automated vehicles" refers to vehicles with these high-level capabilities.

Automated vehicles have been operating on public roadways since 1994, with the field-testing of Daimler-Benz's VITA-2 and Universität der Bundeswehr München's VaMP in Europe *(2)*. The following year, Carnegie Mellon's Navlab 5 vehicle *(3)* demonstrated autonomous steering in the United States *(4)*, while the University of Parma's automated vehicle ARGO completed autonomous trips in 1998 *(5)*. The United States Department of Transportation formed the National Automated Highway System Consortium to prototype fully-automated highway driving, resulting in a 1997 demonstration of vehicle platooning and automated driving aided by

Noah Goodall, Research Scientist, Virginia Center for Transportation Innovation and Research, 530 Edgemont Road, Charlottesville, VA, 22903, noah.goodall@vdot.virginia.gov

embedded magnets and radar-reflective tape in the roadway *(6)*. Automated vehicles from three teams completed the Defense Advanced Research Projects Agency (DARPA) Grand Challenge of 2007 by navigating a complex urban environment within a time limit *(7)*. In 2010, Google announced it had been testing a fleet of seven automated vehicles on public roadways, with over 140,000 miles driven with occasional human intervention *(8)*. Several major automakers have since announced research efforts, including Audi *(9)*, Ford *(10)*, BMW *(11)*, Mercedes-Benz *(12)*, General Motors *(13)*, and Toyota *(14)*.

**TABLE 1 NHTSA Road Vehicle Automation Levels**

| **NHTSA Automation Level** *(1)* | **Description** |
|---|---|
| Level 0: No-Automation | Driver is in complete control over steering, braking, and throttle, although vehicle may provide warnings. |
| Level 1: Function-specific Automation | Vehicle may independently automate one or more control functions. |
| Level 2: Combined Function Automation | At least two control functions are automated and operate in conjunction, for example adaptive cruise control and lane centering. Driver may have to take control with no notice. |
| Level 3: Limited Self-driving Automation | Driver can cede full control to the vehicle in some situations, and driver has a reasonable amount of transition time before he/she must take control. |
| Level 4: Full Self-driving Automation | Vehicle can safely pilot the vehicle for an entire trip, with no expectation for the driver to take control. Such a vehicle does not yet exist. |

### Objective

While there has been a great deal of work in road vehicle automation and obstacle avoidance, to our knowledge there has been no published research in optimal crashing strategies for levels 3 and 4 automated vehicles. Furthermore, of the existing laws governing automated vehicle behavior within the United States, none address computerized control of pre-crash or crash avoidance behavior, and instead require a human be available to take control of the vehicle without notice *(15–17)*. The objective of this study is to assess the need for a moral component to automated vehicle decision making during unavoidable crashes, and to identify the most promising strategies from the field of machine ethics for application in road vehicle automation.

The remainder of this paper focuses on three arguments: that even perfectly-functioning automated vehicles will crash, that certain crashes require the vehicle to make complex ethical decisions, and that there is no obvious way to encode human ethics in computers. Finally, an incremental, hybrid approach for developing ethical automated vehicles is proposed.

## POTENTIAL FOR AUTOMATED VEHICLE CRASHES

Much of the excitement surrounding automated vehicles seems based on the assumption that they will be safer than human drivers. The empirical evidence does not refute this claim: the Google self-driving cars have traveled over 435,000 miles on public roads as of April 2013 *(18)* with only one crash *(19)*, which Google claims occurred while under control of the human driver *(20)*. These mileages did not represent unassisted automated driving, but were tightly supervised by test drivers required to intervene to prevent hazardous situations. Those close to the project have stated that the vehicle can travel 50,000 miles on freeways without intervention *(21)*.

Automated vehicles cannot yet, however, claim to be significantly safer than humans. Smith has noted that, using a Poisson distribution and national mileage and crash estimates, an automated vehicle would need to drive 725,000 miles on representative roadways without incident and without human assistance to say with 99% confidence that they crash less frequently than a human driver, and 300 million miles if considering only fatal crashes *(22)*. An automated heavy truck, given that today's trucks employ professional, trained drivers and a great portion of their mileage are on relatively safe freeways, would need to travel 2.6 million miles without crashing to demonstrate safety benefits compared to a human driver at 99% confidence, and 241 million miles without a fatality. An automated vehicle has yet to travel these distances unassisted. Relevant calculations are summarized in Table 2.

**TABLE 2 Required Mileage of Automated Vehicles to Demonstrate Safety Benefits**

| | All Vehicles | | Heavy Trucks | |
|---|---|---|---|---|
| | All Crashes | Fatal Crashes | All Crashes | Fatal Crashes |
| Vehicle-miles Traveled (VMT) | $2,954 \times 10^9$ | $2,954 \times 10^9$ | $168 \times 10^9$ | $168 \times 10^9$ |
| No. Vehicles Involved in Crashes | 18,705,600 | 45,435 | 295,900 | 3,200 |
| VMT per Crash | 160,000 | 65,000,000 | 567,000 | 52,000,000 |
| Crash-less VMT Required for Benefit* | 725,000 | 300,000,000 | 2,610,000 | 241,000,000 |

*Poisson distribution, P-value < 0.01, based on 2009 data. Sources: *(22–25)*

Crashes may occur due to software or hardware failures, e.g. losing steering control while on curve. An automated vehicle sold to the public would likely require multiple redundancies, extensive testing, and frequent mandatory maintenance to minimize these types of failures. While any engineering system can fail, it is important to distinguish that, for automated vehicles, even a perfectly-functioning system cannot avoid every collision.

Theoretical research robotics confirms that a crash-free environment is unrealistic. While there are many techniques in the literature for collision avoidance in a dynamic environment *(26)*, most are unable to ensure collision avoidance with unpredictable objects. In response, Fraichard and Asama introduced the concept of *inevitable collision states* for mobile robots, defined as "a state for which, no matter what the future trajectory of the system is, a collision with an obstacle eventually occurs" *(27)*. To ensure a mobile robot's safety, its future trajectories are checked to ensure that at no point it enters an inevitable collision state, e.g. moving towards an obstacle without adequate time to brake. In applying this concept to public roadways, the authors acknowledge that even with perfect sensing, it is impossible to guarantee safety in the presence of unpredictable obstacles such as human-driven vehicles, pedestrians, and wildlife *(28)*. Instead, a probability model of expected vehicle behavior is used to ensure automated vehicles avoid moderate and high risk situations *(29)*. (As of this writing, Google is attempting to patent this concept for automated vehicles *(30)*). When human-driven vehicles deviate from these models and behave unexpectedly, crashes can occur.

It is not difficult to imagine a scenario where a crash is unavoidable, even for an automated vehicle with complete knowledge of its world and negligible reaction time. For example, an automated vehicle is vulnerable when stopped at a traffic signal, surrounded on all sides by queued vehicles, with a distracted truck driver approaching from behind. It's impossible for the automated vehicle to avoid some type of collision, though it may use evasive maneuvers to minimize the impact.

Automated vehicle proponents often cite statistics showing a large percentage of vehicle crashes are at least partially due to human error *(31)*. This in fact adds to the challenge of automated vehicle makers, as their vehicles will be forced to interact with occasionally dangerous human drivers, not to mention pedestrians, bicyclists, wildlife, and debris, for the foreseeable future.

The safety improvements of automated vehicles over human drivers have not been proven statistically, and even in simulations with perfect sensing, crashes are still possible. Due to the dynamic environment of roadway traffic, close proximity to vehicles with high speed differentials, and limited maneuverability of automated vehicles at high speeds, automated vehicle safety cannot be assured. Crash risk simply cannot be eliminated through more sophisticated algorithms or sensors.

## DECISION MAKING DURING CRASHES

### Shortcomings of Human Drivers

With today's vehicles, drivers in unsafe situations must make decisions to avoid collisions, and if a collision is unavoidable, to crash as safely as possible. These decisions are made quickly and under great stress, with little forethought or planning. Automated vehicles today rely on human drivers to take over if part of the system fails, or if the automated vehicle encounters a situation it does not understand, such as construction. A human is expected to be able to take control at any time and with no notice in Level 2 automated vehicles, and with reasonable notice in Level 3 vehicles *(1)*.

Early research suggests that expecting humans to remain alert may be overly optimistic. In a recent study, participants driving semi-automated vehicles on a closed test track showed significant increases in eccentric head turns and secondary tasks such as reading when compared to a control group, although distraction countermeasures were effective in reducing this behavior somewhat *(32)*. If experienced drivers begin to rely on automation technologies, the next generation of drivers—which will have grown up around automated vehicles—will likely be even less vigilant about monitoring the roadway.

### Adequate Warning

In Level 3 automation, the driver is not required to remain attentive, but must be available to take control of the vehicle within a certain amount of time after receiving an alert. NHTSA's definition of Level 3 automation does not specify what length of time constitutes an adequate warning. The American Association of State Highway and Transportation Officials (AASHTO) recommends a minimum distance to allow a driver to perceive an unusual situation and react, from between 200 and 400 meters at speeds of 100 km/hr, depending on the type of road and maneuver *(33)*. This distance would have to be significantly longer if the object in question is approaching from the opposite direction, up to 800 meters.

AASHTO's recommendations are meant for complex situations often requiring lane changes, such as approaching a toll turnstile, and represent the upper end of required distances. Still, crashes can occur with very little notice, and a human may not be able to assess the situation and make a better decision than a computer which has been continuously monitoring the roadway.

The warning system's sensitivity to perceived risk poses an additional challenge. In the early 2000s, the Virginia Tech Transportation Institute equipped vehicles with many sensors and recording devices to study naturalistic driving behavior of test subjects on public roadways *(34)*. In this study, thresholds were used to determine when a vehicle may have experienced a near-crash event. Certain safe maneuvers appeared identical to near crashes after analyzing data from vehicle sensors. For example, a maneuver known as a flying pass, where a vehicle approaches a stopped queue at high speed and abruptly changes lanes into a dedicated left or right turn lane, often appeared indistinguishable from a near crash. Researchers were forced to analyze video footage of the driver's face for signs of alarm *(34)*.

For automated vehicles, misunderstanding a human driver's intentions in situations such as a flying pass could lead to false alarms where a safe situation is interpreted as dangerous. Over time, this could decrease the driver's vigilance. In a Level 3 automated vehicle, it may be unrealistic to ask a driver to take control of the vehicle with less than a few seconds prior to collision, even though this situation may occur often. For these reasons, it is likely that in a computer will maintain control of an automated vehicle when encountering dangerous situations and during crashes.

## ETHICS OF CRASHING

Human drivers may often make poor decisions during and before crashes. Humans must overcome severe time constraints, limited experience with their vehicles at the limits of handling, and a narrow cone of vision. While today's automated vehicles also have somewhat limited sensing and processing power, the focus of this paper is on advanced vehicles with near-perfect systems. This is to anticipate the criticism that future sensors and algorithms will eliminate all crashes, and therefore obviate vehicle ethics research *(35)*. If even perfect vehicles must occasionally crash, then there will always be a need for some type of ethical decision making system.

These advanced automated vehicles will be able to make pre-crash decisions using sophisticated software and sensors that can accurately detect nearby vehicle trajectories and perform high speed avoidance maneuvers, thereby overcoming many of the limitations experienced by humans. If a crash is unavoidable, a computer can quickly calculate the best way to crash based on combination of safety, likelihood of outcome, and certainty in measurements, much faster and with greater precision than a human. The computer may decide that braking alone is not optimal, since at highway speeds it is often be more effective to combine braking with swerving, or even swerving and accelerating in an evasive maneuver.

One major disadvantage of automated vehicles during crashes is that unlike a human driver who can decide how to crash in real-time, an automated vehicle's decision of how to crash was defined by a programmer *ahead of time*. The automated vehicle can interpret the sensor data and make a decision, but the decision itself is a result of logic developed and coded months or years ago. This is not a problem in cases where a crash can be avoided—the vehicle selects the safest path and proceeds. However if injury cannot be avoided, the automated vehicle must decide how best to crash. This quickly become a moral decision, demonstrated in an example from Marcus *(36)*, modified slightly for this paper. In the example, illustrated in Figure 1, an automated vehicle is traveling on a two-lane bridge when a bus traveling in the opposite direction suddenly veers into its lane. The automated vehicle must decide how to react using whatever logic has been programmed in advance. There are three alternatives:

A. Veer left and off the bridge, guaranteeing a severe one vehicle crash.
B. Crash head-on into the bus, resulting in a moderate two-vehicle crash.
C. Attempt to squeeze pass the bus on the right. If the bus suddenly corrects back towards its own lane—a low-probability event given how far the bus has drifted—a crash is avoided. If the bus does not correct itself—a high-probability event—then a severe two-vehicle crash results. This crash would be a small offset crash, which carries a greater risk of injury than the full frontal collision in alternative B *(37)*.

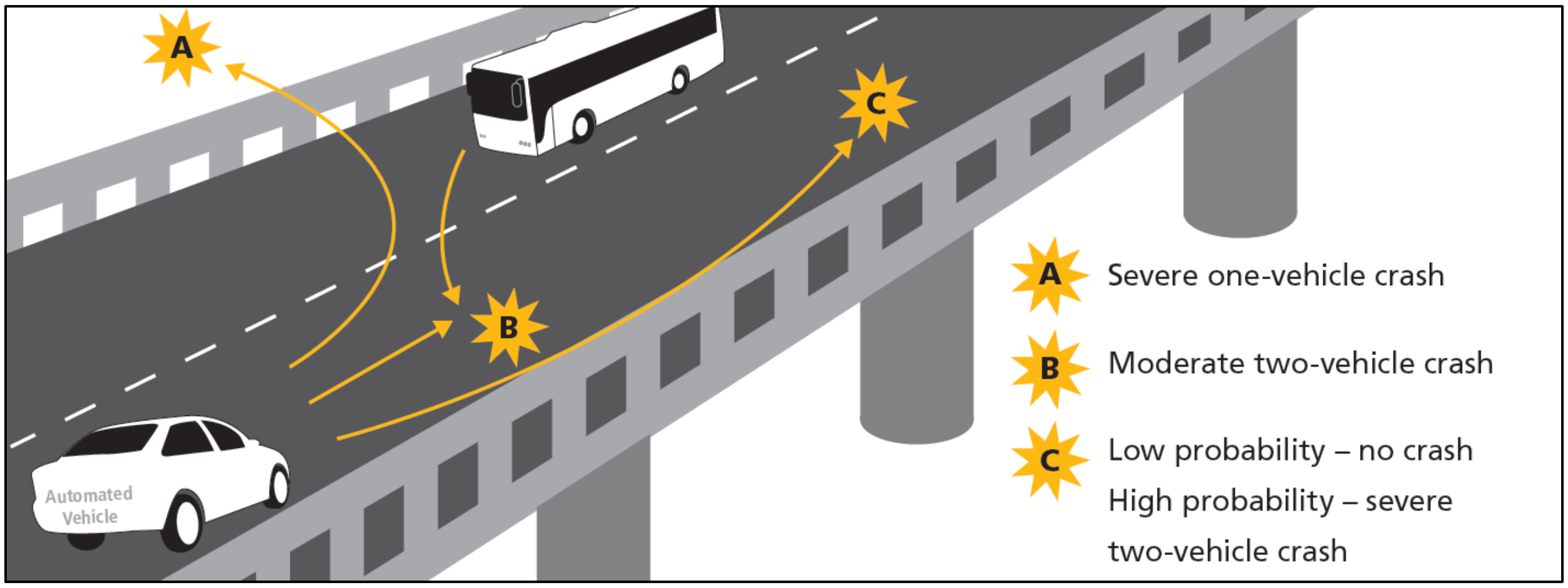

**FIGURE 1 Diagram of three alternative trajectories for an automated vehicle when an oncoming bus suddenly enters its lane.**

It is important to note that these outcomes can only be predicted by the automated vehicle, and are not certain. The automated vehicle's path planning algorithm would have to quickly determine the range of possible outcomes for each considered path, the likelihood of those outcomes occurring, and the algorithm's confidence in these estimates based on quality of sensor data and other factors. Finally, the algorithm would need to somehow optimize an objective function over the range of considered paths and quickly determine the safest route. A primitive form of this objective function could take the following form:

$$f(X) = \sum_{x \in X} s(x)\mathbf{P}(x|X)$$

In the equation, $X$ is the set of possible outcomes $x$ for a single trajectory choice, $s$ is a function representing the severity of a specific outcome $x$, and $\mathbf{P}$ represents the conditional probability of an outcome occurring, given the trajectory choice. The possible outcomes include property damage, injury, and death, each with a range of severities based on additional unlisted factors. Calculating the range of possible choices, outcomes, and probabilities of occurrence are the most technically complex parts of the equation, while calculation of the severity function *s* is the most morally difficult component. Reducing death, injury, and property damage to a comparable number is problematic for many reasons, but without some type of ethical guidance, the automated vehicle has no real way to evaluate risk in the scenario from Figure 1.

## DESIGNING AN ETHICAL VEHICLE

There has been little discussion of the legal and moral implications of automated vehicle decision-making during unavoidable crashes. Most of the research in moral machines have focused on military applications *(38, 39)* or general machine intelligence *(40)*. A relatively recent area of study is *machine ethics,* which focuses on the development of autonomous machines that can exhibit moral behavior when encountering new situations. In this section, an overview of machine ethics and its applications to automated vehicles are discussed.

### Rational Approaches

The instinct for engineers is to directly instruct the automated system how to behave in a variety of circumstances. This rationalist approach often takes the form of either deontology, where the automated system must adhere to a set of rules, or consequentialism, where the system's goal is to maximize some benefit. These rational approaches often appeal to engineers because computers can easily follow rules and maximize functions. Unfortunately, this strategy has several shortcomings, as described using the following examples.

#### *Asimov's Laws of Robotics*

When many first encounter machine ethics, they are reminded of Isaac Asimov's Laws of Robotics *(41),* reprinted here with references to computers replaced by automated vehicles.

1. An automated vehicle may not injure a human being or, through inaction, allow a human being to come to harm.
2. An automated vehicle must obey orders given it by human beings except where such orders would conflict with the First Law.
3. An automated vehicle must protect its own existence as long as such protection does not conflict with the First or Second Law.

Asimov developed these laws, not as a proposal for a rule-based robot code of ethics, but as a literary device to demonstrate the flaws in such a system. Machines are incredibly literal, and any system of ethics with only three rules requires the follower possess some common sense or intuition, particularly if there is a conflict between the rules or within a single rule.

For example, the first rule forbids an automated vehicle to allow a human to "come to harm." A literal interpretation of this rule would prohibit sudden braking, even to avoid a collision, because it would cause whiplash for the vehicle occupants. Wallach and Allen provide an example of a life-saving surgery, where a robot would be unable to "cause harm" and make the first incision *(42)*. A driverless car following these rules may refuse to shift into gear, as leaving the driveway and entering traffic carries some risk of injuring a human.

One could continuously add or clarify rules to cover these unique scenarios. This is a somewhat futile task, as vehicles continue to encounter novel situations, freak accidents, new road designs, and vehicle system upgrades that would need to be addressed. This also requires someone to decide with absolute moral authority what is right and wrong in every roadway situation. In the previous example shown in Figure 1, what if the bus has no passengers? What if it has one, or two, or three passengers? Does it matter if they are adults or children? Ethics as a field has not advanced to this point. As noted by Beavers *(43)*, "after two millennia of moral inquiry, there is still no consensus on how to determine moral right and wrong." Although deontological ethics can provide guidance in many situations, it is not suitable as a complete ethical system because of the incompleteness of any set of rules and the difficulty of articulating complex human ethics as a set of rules.

*Consequentialism*

Asimov's laws may seem too abstract given the science fiction background. A more familiar rational approach for automated vehicle ethics is consequentialism. In this system, a vehicle decides on a crash trajectory that minimizes global harm or damage. If the options are collision vs. no collision, or property damage vs. serious injury, then the choice is obvious.

The way damage is calculated, however, can lead to undesirable outcomes. For example, how does one quantify harm or damage? The most obvious way is in dollars, using damage estimates from the insurance industry. This creates problems in practice. When trying to minimize cost, automated vehicles would choose to collide with the less expensive of two vehicles when given a choice. If the collision is severe and injury is likely, the automated vehicle would choose to collide with the vehicle with the higher safety rating, or choose to collide with a helmeted motorcyclist instead of a helmet-less rider. Many would consider this unfair, not only because of discrimination but also because those who paid for safety are targeted while those who did not are spared. The fact that these decisions were made by a computer algorithm is no consolation.

A second problem with consequentialism is determining what information to incorporate in the decision and what to leave out. For example, crash compatibility measures the damage inflicted by one type of vehicle in collisions with another type *(44)*. Automated vehicles could detect nearby vehicle types from its vision system and, in an unavoidable collision, attempt to collide with more compatible vehicles. Although this may improve safety, it is morally unacceptable, as safer vehicles are unfairly targeted.

Even if the two vehicles in a collision are identical, occupants experience different risks based on demographics. Leonard Evans summarizes an analysis of the Fatality Analysis Reporting System (FARS) data set *(45)*:

> "If one driver is a man, and the other a similar-age woman, the woman is 28% more likely to die. If one driver is age 20 and the other age 70, the older driver is three times as likely to die. If one driver is drunk and the other sober, the drunk is twice as likely to die (because alcohol affects many body organs, not just the brain) *(46)*. If one driver is traveling alone while the other has a passenger, the lone driver is 14% more likely to die than the accompanied driver, because the accompanied driver is in a vehicle heavier by the mass of its passenger *(47)*."

An automated vehicle using facial recognition technology could estimate the gender, age, or level of inebriation of nearby drivers and passengers, and adjust its objective function accordingly, although, again, many would probably find this disturbing.

Both rational approaches to automated vehicle ethics discussed in this section demonstrate shortcomings. Deontological approaches require abstract rules that may not be applicable or may conflict in specific situations, while consequentialist approaches are so rigid that they produce actions that many consider reprehensible. These undesirable outcomes are due to computers' inherent literalness and humans' inability to articulate our own morals. For these reasons, rational approaches alone have limited applicability for automated vehicle ethics.

**Artificial Intelligence Approaches**

For years, automated language translation relied on rules developed by experts. The expectation was that language could be defined by rules, given enough time to learn the rules and write them out. An alternative approach using algorithms that study and learn language automatically, without formal rules, experienced much more success than rule-based methods *(48)*. These techniques are known as artificial intelligence. Language translation is a fitting analogy for ethical systems—in both areas, artificial intelligence methods are useful when we cannot articulate the rules.

Artificial intelligence methods also have potential to learn human ethics by observing human actions or through rewards for its own moral behavior. The computer can identify the components of ethics on its own, without the need for a human to articulate precisely why an action is or is not ethical. Wallach and Allen refer to these techniques as "bottom-up" approaches, which can include such techniques as genetic algorithms, connectionsim, and learning algorithms *(42)*. Artificial neural networks, which use layers of nodes in a connectionist computing approach to find complex relationships between inputs and outputs, have been used in a simple case to classify hypothetical decisions as either moral or amoral *(49)*. Hibbard, in formulating a consequentialist approach to machine ethics, proposed a similar method by which an independent artificial intelligence agent calculated the moral weights assigned by humans by polling participants across a wide range of hypothetical situations *(50)*. Interestingly, an early automated vehicle project, Carnegie Mellon's Autonomous Land Vehicle in a Neural Net (ALVINN), used a simple backpropogation-trained artificial neural network to teach itself to steer by watching a human driver for just two minutes *(51)*. A similar technique could be used, with much more training data, to understand how humans choose to behave—or *should* behave— in morally complex driving situations when time is not a factor. The neural network could be trained on a combination of simulation and recordings of crashes and near-crashes, using human feedback as to the ethical response.

Artificial intelligence techniques have several shortcomings. If not carefully designed, they risk emulating how humans behave rather than what they believe. For example, a human may choose to push a nearby vehicle into oncoming traffic in order to avoid his own collision. Self-preservation instincts which do not maximize overall safety may be realistic, but not ethical. Ethics addresses how humans ought or want to behave, rather than how they actually behave, and artificial intelligence techniques should capture ideal behavior.

Another shortcoming of some artificial intelligence approaches is traceability. Artificial intelligence can be very complex, and artificial neural networks specifically are unable to explain in a comprehensible form how a decision was made based on the input data. Already there are anecdotes of computers discovering relationships in science that researchers do not understand *(52)*. The relationships exist, but are incomprehensible to humans, hidden in gigabytes of data and linkages within an artificial neural network. Bostrom and Yudkowsky have argued that opaque systems cannot be inspected, are unpredictable, and can be easily manipulated *(53)*. They recommend using decision trees to encourage transparency, which is another form type of deontology (Asimov's laws can be easily formulated as a decision flow chart). The risk of manipulation, however, is of particular importance in road vehicle automation. Ethics would likely require that all humans be given equal value, yet a vehicle manufacturer has an incentive to build vehicles that protect its own occupants foremost. Who would buy a car that might protect a stranger at the risk of you or your family's safety? Building a self-protection component into the automated vehicle's ethics could be hidden in a complicated neural network, and would only be discoverable when analyzing long term crash trends. Safeguards must be in place to ensure this does not happen.

While artificial intelligence approaches allow computers to learn human ethics without requiring humans to perform the difficult task articulating ethics as code, they produce actions which cannot be justified or explained in an understandable way. If trained with a narrow set of data, an artificial intelligence may learn behaviors which were completely unintended and undesirable. Without further testing, artificial intelligence approaches cannot be recommended for automated without human-designed rules to increase transparency and prevent obviously unethical behavior.

## PROPOSED ETHICAL VEHICLE DEPLOYMENT STRATEGY

This study investigated issues in ethical decision making in automated vehicles from findings in philosophy, artificial intelligence, and robotics. Based on the identified strengths and weakness of both rational and artificial intelligence approaches, the following three-phase approach is proposed, to be enforced as technology become available.

### Phase 1: Rational Ethics

The first phase, feasible using current technology, would use a rational system for automated vehicle ethics. This system would reward behaviors that minimize global damage. The standards for such a system should be agreed upon by developers of automated vehicles, lawyers, transportation engineers, and ethicists, and should be open and transparent to discourage automakers from building in excessive self-protection into the algorithms. Rules should consist of widely agreed upon concepts, e.g. injuries are preferable to death, property damage is preferable to injury, and vulnerable users should be protected foremost.

A safety metric should be developed to be used in situations where the higher level rules do not specify a behavior, for example when two alternatives both result in similar injuries. This safety metric should be independent of current insurance industry standards, and instead utilize expertise from ethicists and from existing research. A possible starting point for such a system would be value-of-life estimates used in medicine and determining organ transplant recipients, where a complex moral decision must have a numerical basis.

It is unlikely that any human-developed rule set governing robotic behavior will cover all possible scenarios. In any scenarios not covered using these rules, or where the rules conflict or the ethical action is uncertain, the vehicle should brake and evade.

### Phase 2: Hybrid Rational and Artificial Intelligence Approach

In the second phase, which requires sophisticated software which does not yet exist, an automated vehicle's software can use machine learning techniques to understand the correct ethical decision, while bound by the rule-based system in Phase 1. A neural network is a likely candidate method for this approach. The neural network would be trained on a combination of simulation and recordings of crashes and near-crashes. Humans would score potential actions and results as more or less ethical, and would be allowed to score outcomes without the time constraint of an actual crash.

Similar concepts have been promoted. Powers has argued for *adaptive incrementalism in machine ethics*, which while not specifying the technique used to develop an ethical system, acknowledges that a rational-comprehensive approach is impractical for machine ethics *(54)*, as it restricts progress due to limited knowledge in ethics and computer engineering. An incremental approach, where a computer can train itself, would allow progress in automated vehicle research in the absence of a perfect moral system. Wallach and Allen described the limitations of both top-down and bottom-up approaches, and recommended a similar hybrid approach *(42)*.

One must be careful to provide the algorithm with a diverse set of training scenarios. If given a training set that is too narrow in scope, the algorithm can learn morals that were not intended, similar to an extremist. To ensure reasonable behavior of the vehicle, boundaries should be provided. In Phase 2, the rule system from Phase 1 should remain in place as boundary conditions, and the machine learning approach should focus on scenarios not covered by the Phase 1 rules.

**Phase 3: Feedback using Natural Language**

Of the available artificial intelligence methods, artificial neural networks are well suited for classifying the morality of a vehicle's decision due to their high classification accuracy using large volumes of data with low computational costs. A major shortcoming of a neural network is its inability to explain its decision. Unlike a decision tree, where the logic can be traced back over several steps to its source, a neural network is not easily reverse engineered, and it can be very difficult to determine how it arrived at its decision. In an automated vehicle crash, understanding the logic behind an automated vehicle's actions will be critical, particularly if the vehicle did not behave as expected. The use of on-board data collection may allow researchers to recreate a vehicle’s behavior; but even with extensive testing, there can only be a probabilistic understanding of its mechanisms in any situation. Without knowing why an automated vehicle behaves a certain way, there is no way to fix the problem to ensure that it won't happen again.

In order to improve neural network's comprehensibility, computer scientists have developed techniques to extract rule-based explanations from neural networks that are understandable by a human. This process essentially translates a neural network's internal knowledge into a set of symbolic rules, which can then be expressed as natural language *(55)*. Not every decision is accurately represented by a rule, and some rules may be overly complex. Regardless, rule extraction will likely be a useful starting point to understanding the logic of neural networks, and similarly the decisions of automated vehicles.

While recent research shows promise *(56)*, rule extraction is currently in the research phase. As the science progresses, it should be implemented into automated vehicle ethics systems which use the artificial intelligence techniques from Phase 2.

**Summary of Three-Phase Approach**

The three phases constitute an incremental approach to automated vehicle ethics. Its best analogy may be the moral education of a child, discussed by Wallach and Allen *(42)*. Although a child does not have his full moral ability, and may never reach the highest stage of moral development, parents still enforce behavioral boundaries. Parents also encourage the child to consider ways of thinking about morals, in the hopes that they will one day reach a higher stage. The three-phase approach is essentially an attempt to teach a computer ethics, while ensuring it behaves ethically while it learns.

## CONCLUSIONS

Three arguments were made in this paper: automated vehicles will almost certainly crash, even in ideal conditions; an automated vehicle's decisions preceding certain crashes will have a moral component; and there is no obvious way to effectively encode human morality in software.

A three-phase strategy for developing and regulating moral behavior in automated vehicles was proposed, to be implemented as technology progresses. The first phase is a rationalistic moral system for automated vehicles that will take action to minimize the impact of a crash based on generally agreed upon principles, e.g. injuries are preferable to fatalities. The second phase introduces machine learning techniques to study human decisions across a range of real-world and simulated crash scenarios to develop similar values. The rules from the first approach remain in place as behavioral boundaries. The final phase requires an automated vehicle to express its decisions using natural language, so that its highly complex and potentially incomprehensible-to-humans logic may be understood and corrected.

Researchers have made incredible advances in road vehicle automation, with potentially immense safety benefits. Many of the problems faced by automated vehicles can be overcome with better algorithms or sensors. In contrast, the ethical decision making of automated vehicles requires that a vehicle not only behave ethically, but also to understand and apply ethics in new situations, even when humans are not in agreement as to the ethical choice. Further research into machine ethics is encouraged as it applies to road vehicle automation, particularly in the ability of existing crash mitigation systems to behave ethically in realistic scenarios, the types and frequencies of roadway situations requiring ethics, and the legitimacy of a vehicle's tendency to protect is passengers foremost. States beginning to legislate vehicle automation should consider not only the general pre-crash behavior of these vehicles, but also the logic and "values" these vehicles possess.

## ACKNOWLEDGEMENTS

Special thanks to Barbara Johnson of the University of Virginia and Patrick Lin of the California Polytechnic State University for reviewing drafts of this paper.

## REFERENCES

1. National Highway Traffic Safety Administration. *Preliminary Statement of Policy Concerning Automated Vehicles*. Publication NHTSA 14-13. National Highway Traffic Safety Administration, Washington DC, May 2013.
2. Dickmanns, E. D. The Development of Machine Vision for Road Vehicles in the Last Decade. *Proceedings of the IEEE Intelligent Vehicle Symposium,* Vol. 1, June 2002, pp. 268–281.
3. Jochem, T., D. Pomerleau, B. Kumar, and J. Armstrong. PANS: A Portable Navigation Platform. *Proceedings of the Intelligent Vehicles '95 Symposium,* Detroit, MI, Sep. 1995.
4. Pomerleau, D. RALPH: Rapidly Adapting Lateral Position Handler. *Proceedings of the Intelligent Vehicles '95 Symposium,* Detroit, MI, Sep. 1995.
5. Broggi, A., M. Bertozzi, and A. Fascioli. Architectural Issues on Vision-Based Automatic Vehicle Guidance: The Experience of the ARGO Project. *Real-Time Imaging*, Vol. 6, No. 4, Aug. 2000, pp. 313–324.

6. Transportation Research Board. *National Automated Highway System Research Program: A Review - Special Report 253*. The National Academies Press, Washington, DC, 1998.
7. Markoff, J. Crashes and Traffic Jams in Military Test of Robotic Vehicles. *The New York Times*, Nov 5, 2007.
8. Markoff, J. Google Cars Drive Themselves, in Traffic. *The New York Times*, Oct 9, 2010.
9. Hachm, M. CES 2013: Audi Demonstrates Its Self-Driving Car. *Popular Science*, Jan. 9, 2013. http://www.popsci.com/cars/article/2013-01/ces-2013-audi-demonstrates-its-self-driving-car. Accessed June 27, 2013.
10. Newcomb, D. Ford Inches Toward Autonomous Cars, Helps the Parking-Challenged. *Wired: Autopia*, June 26, 2012. http://www.wired.com/autopia/2012/06/ford-tech-driving-challenged/. Accessed June 27, 2013.
11. Travers, J. BMW Traffic Jam Assistant Puts Self-Driving Car Closer Than You Think. *Consumer Reports*, June 11, 2013. http://news.consumerreports.org/cars/2013/06/bmw-traffic-jam-assistant-puts-self-driving-car-closer-than-you-think.html. Accessed June 27, 2013.
12. Daimler. *The new Mercedes-Benz S-Class - Intelligent Drive: Networked with All Senses.* Undated. https://www.daimler.com/dccom/0-5-1597521-1-1597533-1-0-0-1597522-0-0-135-0-0-0-0-0-0-0-0.html. Accessed June 27, 2013.
13. Newman, J. Cadillac Has Self-Driving Cars, Too. *Time*, Apr. 20, 2012. http://techland.time.com/2012/04/20/cadillac-has-self-driving-cars-too/. Accessed June 27, 2013.
14. Guizzo, E. Toyota's Semi-Autonomous Car Will Keep You Safe. *IEEE Spectrum - Automation*, Jan. 8, 2013. http://spectrum.ieee.org/automaton/robotics/artificial-intelligence/toyota-semi-autonomous-lexus-car-will-keep-you-safe. Accessed June 27, 2013.
15. *Vehicles with Autonomous Technology.* Florida Senate Bill CS/HB 1207. 2012.
16. *Vehicles: Autonomous Vehicles: Safety Requirements.* Florida Senate Bill SB 1298. 2012.
17. *Autonomous Vehicle Act of 2012.* Council of the District of Columbia Bill B19-0913. 2012.
18. T.S. How does a self-driving car work? *The Economist*, April 29, 2013. http://www.economist.com/blogs/economist-explains/2013/04/economist-explains-how-self-driving-car-works-driverless. Accessed Aug. 1, 2013.
19. Hyde, J. This is Google's First Self-Driving Car Crash. *Jalopnik*, Aug. 5, 2011. http://jalopnik.com/5828101/this-is-googles-first-self+driving-car-crash. Accessed June 18, 2013.
20. Yarrow, J. Human Driver Crashes Google's Self Driving Car. *Business Insider,* Aug. 5, 2011. http://www.businessinsider.com/googles-self-driving-cars-get-in-their-first-accident-2011-8. Accessed June 19, 2013.
21. Bilger, B. Auto Correct: Has the Self-driving Car at Last Arrived? *The New Yorker,* Nov. 25, 2013.
22. Smith, B. W. Driving at Perfection. The Center for Internet and Society at Stanford Law School, March 2012. http://cyberlaw.stanford.edu/blog/2012/03/driving-perfection. Accessed Oct. 3, 2012.
23. United States Census Bureau. *Statistical Abstract of the United States*. Publication Table 1107. Vehicles Involved in Crashes by Vehicle Type, Rollover Occurrence, and Crash Severity: 2009. United States Census Bureau, Washington, DC, 2012.
24. Office of Freight Management and Operations. *Freight Facts and Figures 2011*. Publication FHWA-HOP-12-002. Federal Highway Administration, Washington, DC, Nov. 2011.

25. National Highway Traffic Safety Administration. *Traffic Safety Facts 2009: A Compilation of Motor Vehicle Crash Data from the Fatality Analysis Reporting System and the General Estimates System*. Publication DOT HS 811 402. National Highway Traffic Safety Administration, Washington, DC, 2009.
26. Laugier, C., and R. Chatila, Eds. *Autonomous Navigation in Dynamic Environments*. Springer-Verlag Berlin Heidelberg, Berlin, Heidelberg, 2007.
27. Fraichard, T., and H. Asama. Inevitable Collision States - A Step Towards Safer Robots? *Advanced Robotics*, Vol. 18, No. 10, 2004, pp. 1001–1024.
28. Benenson, R., T. Fraichard, and M. Parent. Achievable Safety of Driverless Ground Vehicles. *Proceedings of the 10th International Conference on Control, Automation, Robotics and Vision (ICARCV)*, 2008, pp. 515-521.
29. Bautin, A., L. Martinez-Gomez, and T. Fraichard. Inevitable Collision States: A Probabilistic Perspective. *Proceedings of the 2010 IEEE International Conference on Robotics and Automation (ICRA)*, 2010, pp. 4022-4027.
30. Ferguson, D. I. F., and D. A. Dolgov. *Modifying Behavior of Autonomous Vehicle Based on Predicted Behavior of Other Vehicles*, United States Patent Application 20130261872 Kind Code: A1, Oct. 3, 2013.
31. Rumar, K. The Role of Perceptual and Cognitive Filters in Observed Behavior. In *Human Behavior and Traffic Safety* (L. Evans and R. C. Schwing, eds.), Springer US, pp. 151–170.
32. Llaneras, R. E., J. A. Salinger, and C. A. Green. Human Factors Issues Associated with Limited Ability Autonomous Driving Systems: Drivers' Allocation of Visual Attention to the Forward Roadway. *Proceedings of the Seventh International Driving Symposium on Human Factors in Driver Assessment, Training, and Vehicle Design*, Bolton Landing, NY, June, 2013.
33. American Association of State Highway and Transportation Officials. *A Policy on Geometric Design of Highways and Streets*. AASHTO, Washington, DC, 2011.
34. Dingus, T. A., S. G. Klauer, V. L. Neale, A. Petersen, S. E. Lee, J. Sudweeks, M. A. Perez, J. Hankey, D. Ramsey, S. Gupta, C. Bucher, Z. R. Doerzaph, J. Jermeland, and R. R. Knipling. *The 100-Car Naturalistic Driving Study, Phase II - Results of the 100-Car Field Experiment.* Publication DOT HS 810 593. Virginia Tech Transportation Institute, Apr. 2006.
35. Goodall, N. J. Machine Ethics and Automated Vehicles. In *Road Vehicle Automation* (G. Meyer and S. A. Beiker, eds.), Springer, Berlin, 2014.
36. Marcus, G. Moral Machines. *The New Yorker Blogs*, Nov. 27, 2012. http://www.newyorker.com/online/blogs/newsdesk/2012/11/google-driverless-car-morality.html. Accessed Mar. 8, 2013.
37. Sherwood, C. P., J. M. Nolan, and D. S. Zuby. Characteristics of Small Overlap Crashes. *Proceedings of the 21st (ESV) International Technical Conference on the Enhanced Safety of Vehicles*, Stuttgart, Germany, 2009.
38. Arkin, R. C. *Governing Lethal Behavior in Autonomous Robots*. CRC Press, Boca Raton, 2009.
39. Finn, A., and S. Scheding. *Developments and Challenges for Autonomous Unmanned Vehicles: A Compendium*. Springer, Berlin, 2010.
40. Meuhlhauser, L., and L. Helm. Intelligence Explosion and Machine Ethics. In *Singularity Hypotheses: A Scientific and Philosophical Assessment*, Springer, 2012, pp. 101–126.
41. Asimov, I. Runaround. *Astounding Science Fiction*, Mar, 1942, pp. 94–103.

42. Wallach, W., and C. Allen. *Moral Machines: Teaching Robots Right from Wrong*. Oxford University Press, Oxford ; New York, 2009.
43. Beavers, A. F. Moral Machines and the Threat of Ethical Nihilism. In *Robot Ethics: The Ethical and Social Implication on Robotics*, MIT Press, Cambridge, MA, 2011.
44. Summers, S., A. Prasad, and W. T. Hollowell. *NHTSA's Vehicle Compatibility Research Program*. Publication 1999-01-0071. SAE International, Warrendale, PA, Mar. 1999.
45. Evans, L. Death in Traffic: Why Are the Ethical Issues Ignored? *Studies in Ethics, Law, and Technology*, Vol. 2, No. 1, Apr. 2008.
46. Evans, L. *Traffic Safety*. Science Serving Society, Bloomfield, MI, 2004.
47. Evans, L. Causal Influence of Car Mass and Size on Driver Fatality Risk. *American Journal of Public Health*, Vol. 91, No. 7, Jul. 2001, pp. 1076–1081.
48. Russell, S. J., and P. Norvig. *Artificial Intelligence: A Modern Approach*. Prentice-Hall, Upper Saddle River, NJ, 2010.
49. Guarini, M. Particularism and the Classification and Reclassification of Moral Cases. *IEEE Intelligent Systems*, Vol. 21, No. 4, 2006, pp. 22–28.
50. Hibbard, B. Avoiding Unintended AI Behaviors. *Proceedings of the Artificial General Intelligence: 5th International Conference,* Oxford, UK, 2012.
51. Batavia, P., D. Pomerleau, and C. Thorpe. *Applying Advanced Learning Algorithms to ALVINN*. Publication CMU-RI-TR-96-31. Robotics Institute, Carnegie Mellon University, Pittsburgh, PA, Oct. 1996.
52. Arbesman, S. Explain It to Me Again, Computer. *Slate Magazine,* Feb. 25, 2013. http://www.slate.com/articles/technology/future_tense/2013/02/will_computers_eventually_make_scientific_discoveries_we_can_t_comprehend.single.html. Accessed Feb. 25, 2013.
53. Bostrom, N., and E. Yudkowsky. The Ethics of Artificial Intelligence. In *Cambridge Handbook of Artificial Intelligence*, Cambridge University Press, 2013.
54. Powers, T. M. Incremental Machine Ethics. *IEEE Robotics Automation Magazine,* Vol. 18, No. 1, 2011, pp. 51–58.
55. Tickle, A. B., R. Andrews, M. Golea, and J. Diederich. The Truth Will Come to Light: Directions and Challenges in Extracting the Knowledge Embedded within Trained Artificial Neural Networks. *IEEE Transactions on Neural Networks,* Vol. 9, No. 6, 1998, pp. 1057–1068.
56. Augasta, M. G., and T. Kathirvalavakumar. Reverse Engineering the Neural Networks for Rule Extraction in Classification Problems. *Neural Processing Letters,* Vol. 35, No. 2, Apr. 2012, pp. 131–150.